\documentstyle[preprint,aps]{revtex}
\tighten
\begin{document}
\draft

\title{Fermi Edge Singularities and Backscattering \\
in a Weakly Interacting 1D Electron Gas}

\author{C. L. Kane}
\address{Department of Physics, University of Pennsylvania,
Philadelphia, PA 19104}
\author{K. A. Matveev,$^*$ and L. I. Glazman}
\address{Theoretical Physics Institute, University of Minnesota, Minneapolis,
MN
55455}
\maketitle

\begin{abstract}
The photon-absorption edge in a weakly interacting one-dimensional electron
gas is studied, treating backscattering of conduction electrons from the
core hole exactly.
Close to threshold, there is a power-law singularity in the absorption,
$I(\epsilon) \propto \epsilon^{-\alpha}$, with
$\alpha = 3/8 + \delta_+/\pi - \delta_+^2/2\pi^2$ where $\delta_+$ is
the forward scattering phase shift of the core hole.  In contrast to
previous theories, $\alpha$ is finite (and universal)
in the limit of weak core hole potential.
In the case of weak backscattering $U(2k_F)$, the
exponent in the power-law dependence of absorption on energy
crosses over to a value $\alpha = \delta_+/\pi - \delta_+^2/2\pi^2$
above an energy scale $\epsilon^* \sim
[U(2k_F)]^{1/\gamma}$, where $\gamma$ is a dimensionless measure of the
electron-electron interactions.
\end{abstract}

\pacs{PACS numbers: 78.70.Dm, 73.20.Dx}
\narrowtext

The understanding of the nature of the singularities in the
X-ray-absorption edge in metals has played an important role in modern
condensed matter physics\cite{Mahan,Nozieres}.   With the advent of new
microelectronics technology it has become possible to study the related
Fermi edge singularities in one-dimensional (1D) quantum
wires\cite{Calleja}.   Since in one dimension electron-electron interactions
destroy the Fermi surface, it is  an important problem to understand how
the combined effects of  reduced dimensionality and interactions affect the
qualitative nature of the edge singularities.

In 3D metals, the power-law singularity in the absorption edge is
determined
by the famous relation,
\begin{equation}
I(\epsilon) \propto \left(\frac{D_0}{\epsilon}\right)^\alpha
\theta(\epsilon)
\label{0}
\end{equation}
with the exponent $\alpha$ is given by
\begin{equation}
\alpha = 2\delta_0/\pi - \sum_l (\delta_l/\pi)^2.
\label{1}
\end{equation}
Here $\epsilon=\hbar(\omega - \omega_{\rm th})$ is the energy of a
photoelectron counted from the Fermi level, $\omega_{\rm th}$ is the
absorption threshold frequency, $D_0$ is the conduction electron
bandwidth, and $\delta_l$ are the scattering phase shifts associated with
the core hole seen by electrons at the Fermi energy.  It is assumed that
$l=0$ is the dominant channel.  In a non-interacting 1D electron gas,
there are two phase shifts, $\delta_e$ and $\delta_o$, corresponding to
wave functions which are even and odd about the origin. It is more
convenient to introduce linear combinations of these phase shifts,
$\delta_{\pm} = \delta_e \pm \delta_o$. In this representation, the result
(\ref{1}) has the form,
\begin{equation}
\alpha = \frac{\delta_+ + \delta_-}{\pi} -
\frac{\delta_+^2 + \delta_-^2}{2\pi^2}.
\label{2}
\end{equation}
In the Born approximation, the new phase shifts are related to two
different Fourier components of the core hole potential $U(q)$,
\begin{equation}
\delta_+ = \frac{U(0)}{\hbar v_F},\quad
\delta_- = \frac{U(2k_F)}{\hbar v_F},
\label{2.1}
\end{equation}
where $v_F$ and $k_F$ are the Fermi velocity and wavevector.

Recently, a number of authors\cite{Ogawa,Lee} have addressed the effect of
electron-electron interactions in 1D metal in a simplified model with
$U(2k_F)=\delta_- =0$. They concluded that the power-law structure
survives with an exponent modified by the interactions. However, $\alpha$
remains small if the scattering potential $U(0)$ is weak.

In this paper we show that even a small $2k_F$-scattering changes this
result {\it qualitatively}. In particular, the exponent in Eq. (\ref{0})
always becomes of order of $1$ in the immediate vicinity of the threshold
energy. The growth of $\alpha$ at low energies is caused by the
interaction-induced renormalizations\cite{Apel,Kane} of $2k_F$-scattering
that increase the effective value of $U(2k_F)$ and, correspondingly, of
$\delta_-$. The quantitative behavior of $\delta_-(\epsilon)$ can be found
in the case of weakly interacting electrons, when it is possible to define
the transmission amplitude $t$ and to relate it to the phase shifts by a
standard scattering theory formula,
\begin{equation}
t = e^{i\delta_+} \cos \delta_-.
\label{3}
\end{equation}
In the limit of weak interaction, the renormalized transmission amplitude
can be found\cite{Glazman} at any energy,
\begin{equation}
 t(\epsilon) = \frac{t_0 (\epsilon/D_0)^\gamma}
                    {\sqrt{|r_0|^2+|t_0|^2(\epsilon/D_0)^{2\gamma}}},
\label{4}
\end{equation}
where $|r_0|^2=1-|t_0|^2$ is the bare reflection coefficient, and
$\gamma\equiv[V(0)-V(2k_F)]/2\pi\hbar v_F$ is determined by the Fourier
components $V(q)$ of the electron-electron interaction potential.
Comparing (\ref{3}) with (\ref{4}) we conclude that $\delta_+$ does not
depend on energy and corresponds to the unrenormalized value of the
transmission amplitude $t_0$. In agreement with\cite{Kane}, transmission
amplitude vanishes at low energies which implies saturation of the other
phase shift at $\delta_-=\pi/2$. Consequently, according to Eq. (\ref{2}),
very close to the threshold the exponent $\alpha$ is given by
\begin{equation}
\alpha = \frac{3}{8} + \frac{\delta_+}{\pi} -
\frac{\delta_+^2}{2\pi^2}.
\label{5}
\end{equation}
Thus, $\alpha$ near the threshold differs from the results
of works\cite{Ogawa,Lee} by $3/8$. The region of energies where
Eq. (\ref{5}) is valid depends on $U(2k_F)$. For a weak backscattering,
there is a clear crossover between the values given by (\ref{5}) and the
results of\cite{Ogawa,Lee}, as shown in the Figure.
This crossover occurs at the energy
scale $\epsilon^*$ where the phase $\delta_-$ becomes of the order of one.
Using Eqs. (\ref{3}) and (\ref{4}), we find
\begin{equation}
\epsilon^* \sim D_0\left|\frac{r_0}{t_0}\right|^{1/\gamma}
\sim D_0\left|\frac{U(2k_F)}{\hbar v_F}\right|^{1/\gamma}.
\label{6}
\end{equation}
The simple picture presented above has two drawbacks. It uses Eqs.
(\ref{0}) and (\ref{3}) that are valid, strictly speaking, for
non-interacting electrons only. Besides, Eq. (\ref{1}) assumes
energy-independent phase shifts.

In order to establish the above results, it is necessary to develop an
approach that treats both the renormalizations of $\delta_-$ and the
absorption intensity $I(\epsilon)$ in a unified manner. For non-interacting
electrons, a number of techniques have been used  to this end. Nozieres
and de Dominicis\cite{Nozieres} summed the perturbation series in the
strength of the potential. Ohtaka and Tanabe used a technique based on
Slater determinants\cite{Ohtaka}. Schotte and Schotte
employed a bosonization  technique\cite{Schotte}.
The latter approach is the most natural to
generalize to include electron-electron interactions.

Bosonization has proved a very useful technique for studying the
interacting 1D electron gas, or Luttinger liquid
\cite{Bosonization}.
The advantage of bosonization, is that in the absence of backscattering,
the low energy behavior is determined by a free field theory, which
describes the collective density fluctuations,
and exact results for the behavior of correllation functions
may be simply obtained.  Unfortunately, backscattering of electrons
will introduce a nonlinear term into the theory
which cannot be treated exactly.
Schotte and Schotte
employed a different bosonization scheme, however,
which allowed them to treat exactly the scattering of 3D non-interacting
electrons on a core hole\cite{Schotte}. They applied the
bosonization technique to each angular  momentum channel independently.
We use a similar approach for our 1D problem that allows us to treat the
backscattering potential exactly.

Here we briefly outline this nonstandard bosonization transformation.
For $k>0$, instead of usual left ($L$) and right ($R$) movers described
by wave functions $e^{\pm ikx}$ we introduce even ($e$) and odd ($o$) channels
correspoding to $\cos kx$ and $\sin kx$ respectively.
The electron anihilation operators corresponding to these
states are then related by $\psi_{L,R}(k) =
[\psi_e(k) \pm i \psi_o(k)]/\sqrt{2}$. We then define
bosonic fields $\phi_{e,o}(\tilde x)$ such that
\begin{equation}
\tilde\nabla\phi_{e,o}(\tilde x) =
\sum_{k,q} e^{i q \tilde x} \psi_{e,o}^\dagger(k+q)\psi_{e,o}(k).
\end{equation}
(Note that the coordinate $\tilde x$ should not be confused with
the physical coordinate $x$, since the transformation to even and
odd wavefunctions mixes
$x$ and $-x$.)  We may then, as usual, express the electron creation
operator as,
\begin{equation}
\psi_{e,o}(k) = {1\over\sqrt{2\pi\eta}}
\int_{-\infty}^\infty d\tilde x e^{-ik\tilde x}
\exp[i \phi_{e,o}(\tilde x)],
\end{equation}
where $\eta \approx\hbar v_F/D_0$ is the short distance cutoff.
When expressed in terms of these variables, the Hamiltonian
for non-interacting electrons, including backscattering,
 is quadratic in $\phi_{e,o}$.
We write
\begin{equation}
H = H_0 + H_{\rm barrier},
\label{(5)}
\end{equation}
where
$H_0 = \sum_{k>0} \hbar v_F k [\psi_e^\dagger \psi_e + \psi_o^\dagger
\psi_o]$ is the kinetic energy expressed in the $(e,o)$ basis, and
$H_{\rm barrier} = U(2k_F)(\psi_L^\dagger \psi_R + \psi_R^\dagger \psi_L)
+  U(0) ( \psi_L^\dagger\psi_L + \psi_R^\dagger\psi_R)$ is the potential
due to scattering off of the core hole (operators $\psi_L$ and
$\psi_R$ are taken at $x=0$).
Expressed in terms of the bosonic operators, these may be written,
\begin{equation}
H_0 = {\hbar v_F\over{8\pi}} \int_{-\infty}^\infty d\tilde x
\left\{ [\nabla\phi_+(\tilde x)]^2 + [\nabla\phi_-(\tilde x)]^2\right\}
\end{equation}
and
\begin{equation}
H_{\rm barrier} = {\hbar v_F\over{2\pi}} \left[\delta_+ \nabla \phi_+ (\tilde
x=0) + \delta_- \nabla\phi_- (\tilde x=0)\right],
\end{equation}
where $\delta_\pm$ are related to $U(q)$ by Eq.~(\ref{2.1}), and we have
defined $\phi_\pm = \phi_e \pm \phi_o$.   Since this Hamiltonian
is at most quadratic, it is a simple matter to compute the X-ray
response exactly.

We now add an electron-electron interaction to the Hamiltonian (\ref{(5)}),
and  in the long-wavelength limit write
$H_{\rm int} = V \int dx \rho_L(x) \rho_R(x)$,
where the right and left moving electron densities are
$\rho_{L,R}(x) = \psi_{L,R}^\dagger(x)\psi_{L,R}(x)$\cite{g4}.
Expressed in terms of the boson fields, this takes the form,
\begin{equation}
H_{\rm int} =  V \int_{-\infty}^\infty d\tilde x
\left[ {1\over{16\pi^2}}\tilde\nabla \phi_+(\tilde x)
\tilde\nabla \phi_+(-\tilde x)
-{1\over{4\pi^2\eta^2}}  \sin \phi_-(\tilde x)
 \sin \phi_-(-\tilde x) \right].
\label{(8)}
\end{equation}
Note that the Hamiltonian $H_0+H_{\rm int} + H_{\rm barrier}$
decouples into independent $\phi_+$ and $\phi_-$ channels.
When $\delta_- = 0$, the effect of the barrier is contained
only in the part dependent on $\phi_+$, which is quadratic.  Thus,
as shown in references\cite{Ogawa,Lee},  when $\delta_- = 0$ an exact
solution is possible for arbitrary interaction strength.
When $\delta_-$ is finite,
the interaction term containing non-quadratic pieces
makes a general solution difficult.  Nonetheless, it is
straightforward to expand perturbatively in $\gamma = V/2\pi \hbar v_F$,
in order to show that $\delta_-$ changes drastically the X-ray absorption
edge singularity.

The X-ray response may be determined by computing
$I(\epsilon) \propto {\rm Re}\int_{0}^\infty dt e^{i\epsilon
t/\hbar}A(t)$, with
\begin{equation}
A(t) = \langle \psi_e(x=0,t)\exp\left\{- \frac{i}{\hbar} \int_0^t H_{\rm
barrier}(t') dt'\right\}\psi_e^\dagger(x=0,0)\rangle.
\label{(9)}
\end{equation}
The averaging in Eq.~(\ref{(9)}) is performed over the ground state of
the Hamiltonian $H_0+ H_{\rm int}$.  To the lowest order in $\gamma$
we find the correction to $I(\epsilon)$,
\begin{equation}
I(\epsilon) \propto \theta(\epsilon)
\left({D\over\epsilon}\right)^{\alpha_0}
\left[ 1 -
{\gamma\over{4\pi}} \left( {\delta_-\over\pi} - 1\right) \sin(2\delta_-)
\ln^2 {D\over\epsilon}
\right]
\label{(10)}
\end{equation}
with $\alpha_0$ given by (\ref{2}).  For $\gamma=0$ we recover the exact
result for non-interacting electrons\cite{Mahan,Nozieres,Schotte}. The
first correction in $\gamma$ diverges logarithmically as
$\epsilon\rightarrow 0$\cite{g4}.

In logarithmic problems, a renormalization group analysis often allows one
to extend the results of perturbation theory.  Here we shall employ the
usual program for the renormalization group.  In (\ref{(9)}) we thus divide
$\phi_\pm$ into slow and fast components $\phi_\pm = \phi_\pm^< +
\phi_\pm^>$, where $\phi^>$ is composed of Fourier components with
$e^{-\ell} D < q < D $.  Upon integrating out $\phi^>$ and rescaling space
and time by a factor $e^\ell$, we arrive at an equivalent problem with
renormalized parameters. This procedure can easily be carried out
perturbatively in $\gamma$, and we find that the lowest
order renormalization group flow equation is
\begin{equation}
{d\delta_-\over{d\ell}} = {\gamma\over 2} \sin 2\delta_-.
\label{(11)}
\end{equation}
The related evolution of $A(t)$ is given by
\begin{equation}
{d \ln A(t)\over {d\ell}} =  - {1\over 2}({\delta_+\over\pi}-1)^2
-  {1\over 2}({\delta_-\over\pi}-1)^2.
\label{(12)}
\end{equation}

It may be observed that the perturbation theory result (\ref{(10)})
satisfies this scaling relation.  Eq. (\ref{(11)}) has been obtained
earlier, in a study of the transmission coefficient of a barrier in a
weakly interacting electron gas\cite{Glazman}. It shows that for an
arbitrarily small backscattering, phase shift will grow and at low energy
scales saturate at $\pi/2$.
For $\delta_-$ near $0$ or $\pi/2$, equation (\ref{(11)}) is  equivalent
to the weak interaction limit of the renormalization group flow equations
derived for the weak and strong barrier limits in Ref.\onlinecite{Kane}.

Equation (\ref{(11)}) may simply be solved for the phase shift at any
energy scale,
\begin{equation}
\delta_-(\ell) = \tan^{-1} [e^{\gamma\ell} \tan \delta_-(0)].
\label{(13)}
\end{equation}
We may obtain an expression for the correlation function
$A(t)$ by integrating (\ref{(12)}) down to an energy scale of order
$D e^{-\ell} \approx \hbar/t$,
\begin{equation}
A(t) = \exp \left[ - {1\over 2}\left({\delta_+\over\pi}-1\right)^2 \ln
\frac{D_{0}t}{\hbar} - \frac{1}{2}\int_0^{\ln Dt/\hbar}
\left({{\delta_-(\ell)}\over\pi}-1\right)^2d\ell\right].
\label{20}
\end{equation}
In the weak interaction limit in which we are working, the Fourier
transform of (\ref{20}) may be found by noting that $\delta_-$ is a very
slowly varying function of $\ell$, so we obtain
\begin{equation}
I(\epsilon) \propto \left( \frac{D}{\epsilon} \right)^{
-{1\over 2}\left({\delta_+\over\pi}-1\right)^2 +1}
\exp\left[
-{1\over 2}\int_0^{\ln {D\over{\epsilon}}}
\left({{\delta_-(\ell)}\over\pi}-1\right)^2 d\ell
\right].
\label{(15)}
\end{equation}
Here $\delta_-(\ell)$ is given by Eq.~(\ref{(13)}). At relatively large
energies, one should substitute into (\ref{(15)}) the unrenormalized
backscattering phase shift
$\delta_-(0)$. If $\delta_-(0)$ is small, then
$I(\epsilon)$ becomes a power law with the
exponent (\ref{2}). In the limit of small energies, $\ell\to\infty$, the
phase shift $\delta_-=\pi/2$, and (\ref{(15)}) reduces to a power-law
singularity with an exponent given by Eq.~(\ref{5}). As it follows from
Eqs. (\ref{(13)}) and (\ref{(15)}) there is a clear crossover between
these two regimes, if the initial value of phase $\delta_-$ is small. The
crossover occurs at energy $\epsilon^* \sim D_0[\delta_-(0)]^{1/\gamma}$,
in accordance with (\ref{6}).

Since the Hamiltonian (\ref{(8)}) is decoupled into independent $+$ and
$-$ channels, it is clear that the Fermi edge exponent should be the sum
of two independent terms determined by $\delta_+$ and $\delta_-$ even when
the interactions are not weak.  The $\delta_+$ term was computed in
references \cite{Ogawa,Lee}.   Our solution of this problem in the weak
interaction limit gives us strong indication of how the other term behaves
in the intermediate interaction regime. In this regime, it is known that
in a Luttinger liquid with repulsive interactions the renormalized
backscattering grows as $U(2k_F) \propto (D_0/\epsilon)^{1-g}$,
where $g = (1 + 2\gamma)^{-1/2}$\cite{Kane}. Hence there is a crossover between
the limits of weak and strong backscattering that occurs at $\epsilon^*
\sim D_0[U(2k_F)/\hbar v_F]^{1/(1-g)}$.
Recently, Prokof'ev\cite{Prokofev}  has computed the edge exponent at the
strong backscattering fixed point ($\epsilon \to 0$). His result
agrees with ours in the weak interaction limit. This is compelling
evidence that the crossover physics described in the Figure remains valid
in the intermediate interaction regime
(see also Ref. \onlinecite{Gogolin}).

So far we considered the model of spinless electrons. An advantage of
the perturbation theory in $\gamma$ used here is that it allows a
straightforward generalization to the case of spin-$\frac12$ electrons.
The energy dependence of the transmission coefficient (\ref{4}), and
consequently $\delta_-$, is modified\cite{Glazman}. However the
qualitative picture of the X-ray response behavior remains the same.
As in the spinless case, $\delta_-$ renormalizes from its initial
value to $\pi/2$, which leads to a crossover of the power-law exponent
$\alpha$. The limiting values of $\alpha$ may be found from Eq.~(\ref{1}).
At large energies it gives $\alpha = (\delta_++\delta_-)/\pi -
(\delta_+^2+\delta_-^2)/\pi^2$, while near the threshold one obtains
$\alpha=1/4 + \delta_+/\pi - \delta_+^2/\pi^2$.

In conclusion, we studied the X-ray edge singularity for the weakly
interacting 1D electron gas. We have shown that even a weak
backscattering on the core hole affects this singularity drastically.
It leads to the crossover in the dependence of the absorption-edge
exponent on energy, as shown in the Figure. In the case of a weak core hole
potential, the X-ray exponent near the threshold equals $3/8$ for spinless
fermions and $1/4$ for spin-$\frac12$ electrons.

The authors are grateful to A. O. Gogolin for providing the
preprint\cite{Gogolin} prior to publication. C.L.K. acknowledges the
hospitality of the Theoretical Physics Institute at the University of
Minnesota where this work was started. K.A.M and L.I.G. acknowledge
the hospitality of Aspen Center of Physics. This work was supported
by NSF Grant DMR-9117341.

\begin{figure}
\caption{
A Log-Log plot showing the crossover in the X-ray response obtained
by numerically integrating
equation (21).  We have set $\gamma = 0.6$, which is the correct
order of magnitude for a GaAs quantum wire.
The core hole potential
is characterized by $\delta_+ = 0.1$ and $\delta_-(0) = 0.1$.  The
dashed lines indicate the asymptotic power laws for
$\epsilon \ll \epsilon^*$ and $\epsilon \gg \epsilon^*$, with
$\epsilon^*/D \approx .02$.
}
\label{fig1}
\end{figure}


\begin{references}

\bibitem[*]{*} On leave from the Institute of Solid State Physics,
  Chernogolovka, Moscow distr., 142432, Russia.

\bibitem{Mahan}
G. D. Mahan, {\it Many-particle physics}, 2nd ed., (Plenum, New York,
1990), ch.~8.

\bibitem{Nozieres}
P. Nozieres and C.T. deDominicis,
Phys. Rev. {\bf 178}, 1097 (1969).

\bibitem{Calleja}
J.M. Calleja, et. al., Solid State Commun. {\bf 79}, 911 (1991).

\bibitem{Ogawa}
T. Ogawa, A. Furusaki, N. Nagaosa,
Phys. Rev. Lett. {\bf 68}, 3638 (1992).

\bibitem{Lee}
D.K. Lee, Y. Chen, Phys. Rev. Lett. {\bf 69}, 1399 (1992).

\bibitem{Apel}
W. Apel and T.M. Rice, Phys. Rev. B {\bf 26}, 7063 (1982).

\bibitem{Kane}
C.L. Kane and M.P.A. Fisher,
Phys. Rev. Lett. {\bf 68}, 1220 (1992).

\bibitem{Glazman}
D. Yue, K.A. Matveev and L.I. Glazman
``Conduction of a weakly interacting 1D electron gas through a
single barrier,'' preprint TPI-MINN 93/39-T (1993).

\bibitem{Ohtaka}
K. Ohtaka and Y. Tanabe, Rev. Mod. Phys {\bf 62}, 929 (1990).

\bibitem{Schotte}
K.D. Schotte and U. Schotte, Phys. Rev. {\bf 182}, 479 (1969).

\bibitem{Bosonization}
F.D.M. Haldane, J. Phys. C {\bf 14}, 2585 (1981);
J. Solyom, Advances in Physics {\bf 28}, 201 (1970);
V.J. Emery in {\it Highly Conducting One-Dimensional Solids},
edited by J.T. Devreese (Plenum Press, New York 1979).

\bibitem{g4}
It is also possible to define an additional independent interaction of
the form $W \int dx [\rho_L(x)^2 + \rho_R(x)^2]$.  However, it can
be checked that it gives rise to a correction which is less divergent
than the correction in equation (\ref{(10)}).


\bibitem{Prokofev}
N.V. Prokof'ev, ``Fermi-Edge Singularity with Backscattering in the
Luttinger Liquid Model,'' preprint 1993.

\bibitem{Gogolin}
A. O. Gogolin, ``Time-Dependent Perturbation in Luttinger Liquid,''
preprint 1993.
\end{references}
\end{document}